# Large-area uniform and low-cost dual-mode plasmonic naked-eye colorimetry and SERS sensor with handheld Raman spectrometer

Zhida Xu[1,*], Jing Jiang[1], Xinhao Wang[1], Kevin Han[1], Abid Ameen[1], Ibrahim Khan[1], Te-Wei Chang[1], and Gang Logan Liu[1,*]

[1]Department of Electrical and Computer Engineering, University of Illinois at Urbana-Champaign, Urbana, IL 61801

*Corresponding authors: xuzhida1108@gmail.com and loganliu@illinois.edu

Abstract

We demonstrated a highly-sensitive, wafer-scale, high-uniform plasmonic nano-mushroom substrate based on plastic for naked-eye plasmonic colorimetry and surface-enhanced Raman spectroscopy (SERS). We gave it the name FlexBrite. The dual-mode functionality of FlexBrite allows for label-free qualitative analysis by SERS with the enhancement factor (EF) of $10^8$ and label-free quantitative analysis by naked-eye colorimetry with the sensitivity of 611 nm/RIU. The SERS EF of FlexBrite in wet state was found to be $4.81\times10^8$, 7 times stronger than dry state, making FlexBrite suitable for aqueous environments such as microfluid system. The label-free detection of biotin-streptavidin interaction by both SERS and colorimetry was demonstrated with FlexBrite. Detection of trace amount of the narcotic drug methamphetamine in drinking water by SERS was implemented with a handheld Raman spectrometer and FlexBrite. This plastic-based dual-mode nano-mushroom substrate FlexBrite has the potential to be used as a sensing platform for easy and fast analysis of chemical and biological assay.

Introduction

The increasing demands of medical diagnostics, environmental monitoring, drug screening and food safety control entrust the development of highly selective, sensitive and low-cost sensors with foremost importance. Among the various sensors, optical sensors based on plasmonics are becoming the method of choice in label-free analysis of biomolecular interactions. Recently, due to the advancement of nanofabrication technology, noble metallic nanostructures with unique plasmonic properties like surface plasmon resonance (SPR) and localized surface plasmon resonance (LSPR) reveal their advantages in different applications such as chemical and biological sensing, subwavelength imaging and energy harvesting.[1-3] The principle of the SPR or LSPR sensors is mostly based on measuring the shift of plasmonic resonance peak wavelength induced by the refractive index (RI) change of the environment.[4] For higher sensitivity of RI sensing, larger absolute wavelength shift of resonance peak per unit RI change and higher quality factor (height to width ratio) of the resonance peak are desired. But there is usually a trade-off between the two desired properties. Taking the two most commonly used schemes -- subwavelength grating structure and colloidal metallic nanoparticles for example, subwavelength grating structure with

Kretschmann configuration has been proved for its ultrahigh sensitivity (~$10^6$ nm/RIU) with high quality factor, but the absolute wavelength shift is only few nanometers.[5] Such small wavelength shift requires sophisticated spectrometer for detection. On the other hand, LSPR sensing based on colloidal metallic nanoparticles has much larger wavelength shift which requires less on the equipment, making plasmonic colorimetry by naked-eye possible.[6] However, the colloidal metallic nanoparticles sensing has lower accuracy due to its lower quality factor and less uniformity and reproducibility because the particle size, shape and distribution can affect the LSPR performance dramatically and they are difficult to control precisely [7] Plasmonic sensors for RI sensing are very good at label-free quantitative analysis but not good at qualitative analysis. In contrast, Raman spectroscopy is very good at qualitative identification and analysis with the distinctive vibrational features of molecules. However, the cross-section of Raman scattering is so small (1 in 10 million) that highly-sensitive spectrometer or highly concentrated target analyte is required.[8] Fortunately, surface-enhanced Raman spectroscopy (SERS) can significantly enhance the Raman scattering by $10^6$ to $10^{10}$ order thus greatly improve the sensitivity of Raman spectroscopy, even making the detection of single molecule possible.[8, 9] However, SERS, which reply on the specialized noble metallic nanostructures, has its own limitations such as low-repeatability, non-uniformity, high-cost, complicated fabrication process, low throughput and small area. These drawbacks thwart the wide application of SERS. Especially, the low uniformity and small area hamper the quantitative analysis by SERS.[10] Up to date, there have been a lot of work reported on either SERS or plasmonic sensing but very few have reported sensing with the same device for both SERS and plasmonics, let alone plasmonic colorimetry sensing by naked-eye. Sensing based on the shift of plasmonic resonance peak is very sensitive with the detection of RI change thus it is good at quantitative analysis. On the other hand, SERS is very sensitive at the qualitative detection of chemicals by their distinct Raman spectra while not very good at quantitative analysis. The unsuitability of SERS for quantitative analysis is mostly attributed to low reproducibility, non-uniformity and small area of SERS substrates.[11-13] Nanohole-array structure has been demonstrated for both plasmonic colorimetric RI sensing by naked-eye[14, 15] and SERS[16, 17]. Nanopillar or nanocone structure has been designed with different techniques for plasmonics application, colorimetric sensing and Raman scattering enhancement. For example, periodic nanopillar structure was fabricated with electron-beam (e-beam) lithograph[18], focused ion beam (FIB) milling[19] and laser interference lithography[20]. Some of them have good Raman scattering enhancement and some others demonstrate color change with the periodicity of nanostructure. However, all of them are limited by low throughput and high cost. Nano-imprint tehnique with lower cost produced low yield of nano-pillar substrate.[21] Other low-cost methods such as nano-sphere lithography use polystyrene nano-spheres as etching mask to produce semi-periodic nano-pillars[22]. Anodized aluminum oxide (AAO) technique uses porous AAO as a template to form metal nano-pillar[23] or form polymeric nano-pillar after nano-imprint[24]. Even though those nanopillar or nanocone structures are widely used for SERS, none of them demonstrated colorimetry by naked-eye.

In this work, we demonstrated a dual-mode plasmonic colorimetry and SERS substrate based on plastics. Due to its plastic nature, we named this substrate FlexBrite as it is bendable. Fig.1(a) is a photograph of a piece of FlexBrite. The large-area uniformity of this dual-mode sensor makes quantitative Raman analysis and colorimetric RI sensing possible. A scanning electron microscopy (SEM) image of the nanostructure of FlexBrite is nano-mushroom array, shown in Fig.1(b). A photograph of FlexBrite wafer with the diameter of 4 inches wrapped around a mailing tube and SEM images of nano-mushroom array with different perspective angles are included in Fig. 1S as supplements. The special nanomushroom structure differentiates it from ordinary nanocone or nanopillar structure and has special properties for both plasmonic colorimetry and SERS. As shown in Fig. 1(a), FlexBrite shows a purple color without any liquid on the surface from a rough top-view perspective. If the perspective angle becomes larger, then the purple color will gradually change to blue (45º ~60º), green (60º ~75º) and red (75º ~90º). All the colorimetric sensing images were taken with a rough top-view perspective. Fig. 1(c) and (d) roughly demonstrate the colorimetric property of FlexBrite. Fig. 1(c) demonstrates a water droplet stays on the surface of substrate showing green while the FlexBrite substrate itself looks purple. Fig. 1(d) shows that different liquids with difference RI will present different colors on FlexBrite. Water (n = 1.333) looks green while cedar wood oil (n = 1.518) looks red on the surface of FlexBrite. Compared to our previous work of naked-eye colorimetry by Lycurgus nano-cup array structure (nanoLCA), the colorimetric properties of FlexBrite has two major advantages. One advantage is FlexBrite works better for a reflection mode, which makes the measurement easier and more accurate. The colorimetry of NanoLCA sensor works in transmission mode, in which the detector must be placed on the other side of sample liquid. In the transmission mode, the detector and light source are separated on two sides of the sample, so the transmitted light was dimmed by the sample. The plastic substrate and especially the metal layer with thickness of 80 nm to 90 nm, will lower the accuracy of measurement. Also for this reason, the backside of nanoLCA sensor has to be kept very clean. FlexBrite works in the reflection mode, in which the light source and detector are on the same side of the sample so FlexBrite does not have the problems of dimming and backside cleaning. The other advantage is that FlexBrite has wider color range and higher sensitivity for colorimetry. For air, water and oil respectively, nanoLCA shows green, yellow and red color while FlexBrite shows purple, green and yellow/red. In addition to colorimetry, FlexBrite has SERS property with the enhancement factor of $10^8$ and large-area uniformity with small variation of within 15%. We also found that the SERS of FlexBrite excited by 633 nm laser is 7 time stronger in wet state than in dry state, making FlexBrite suitable for aqueous environment such as microfluids. The 7 times enhancement in wet state was explained by both finite-difference time-domain (FDTD) simulation and reflection measurement. We demonstrated two applications of FlexBrite. One is detection of biotin-streptavidin interaction by both SERS and colorimetry while the other is detection of trace amount of the narcotic drug methamphetamine in drinking water by SERS with a handheld Raman spectrometer and FlexBrite.

Characterization of colorimetry of FlexBrite

In Fig.1 we have roughly demonstrated the colorimetric property of FlexBrite. For a quantitative analysis of colorimetry, we prepared glycerol solutions with different concentrations for different RI. The RI of glycerol solutions with different concentrations at room temperature were looked up in a table provided by Dow Chemical[25]. The liquid was dropped on the surface of FlexBrite then covered with a coverslip. Fig. 2(a) is a series of images of FlexBrite with liquid of different RI. These images were taken in a reflection mode with Olympus BX51 upright fluorescence microscope equipped with 20X objective lens, DP50 digital camera and a mercury lamp. We kept the same camera setting for taking all the images. Exposure time of 30 ms. RGB gain was set to be 1. As the RI of liquid increases from 1.333(water) to glycerol solutions with different concentrations (RI from

1.357 to 1.474) then to 1.518(cedar wood oil), the color of FlexBrite changes from green to yellow then to red. Then we can say that the color of FlexBrite is red-shifted as the RI of liquid increases. Fig.2 (b) and (c) show the averaged RGB and HSV values of the images in Fig.2 (a). From Fig. 2(b), we can see the red (R) channel increases while the green (G) and blue (B) decrease as the RI of liquid increases, which matches the red-shift trend observed by naked-eye in Fig. 2(a). The R and G channels almost follow a linear relationship with RI. From Fig. 2(c), we can see the saturation (S) increases while hue (H) decreases with RI increase but value (V) does not follow a clear trend with RI. As a result, we can use R, G, B or S, H to quantitatively measure the RI of liquid on the surface of FlexBrite. Considering R and G have better linearity, we prefer to use R and G as the indicators of RI of liquid. In the supplementary materials, we include a video as Video S1 to show the color change of FlexBrite as we gradually increased the RI of liquid in a microfluidic channel, in which the FlexBrite is placed at the bottom. In Video S1, we started with nothing but air in the microfluidic channel, followed by water and glycerol solutions with concentrations from low to high to observe the color change of FlexBrite. Fig. 2S in the supplement shows the averaged RGB and HSV values change with RI of liquid with time as the x-axis. To prove the reusability of FlexBrite colorimetry, we alternatively replaced the liquid in the microfluidic channel with water and glycerol solutions and recorded the process as Video S2 in the supplements. Fig. 3 is a series of snapshots of Video S2. As we replaced the liquid alternatively with water and glycerol then water again, the color FlexBrite turned from green to red then to green again. Fig. 3S in the supplements shows the RGB and HSV values as we alternatively change the liquid. We can see the RGB and HSV values also alternatively change with time. Taking the R value in Fig. 3S for example, it suddenly rises after water is replaced by glycerol, corresponding to green turns red in Fig. 3. Then the R value drops back to its previous value after glycerol is replaced with water, corresponding to red turns back to green in Fig. 3.

In addition to the image RGB and HSV analysis of colorimetry of FlexBrite, we also used spectroscopy to characterize the colorimetric property of FlexBrite. Fig. 4(a) shows the reflection spectra from liquids with different RI on its surface. The reflection spectra was taken with the same Olympus BX51 upright microscope and Ocean Optics USB2000+ Fiber Optic Spectrometer. We can see the peak of reflectance spectra was shifted to longer wavelength as the liquid RI increases, corresponding to the red-shift we observed in the images in Fig. 2(a). We plotted the peak wavelengths in Fig. 4(a) versus the liquid RI in Fig. 4(c). We calculated the slope of linear regression of the data in Fig. 4(c) to get the sensitivity of 611 nm/RIU. FlexBrite's high RI sensitivity of 611 nm/RIU enabled naked-eye detection.

FDTD simulation

To investigate the mechanism of FlexBrite's high sensitivity in colorimetry and predict the optimal structure of the nano-mushroom, we use finite-difference time-domain (FDTD) method to simulate the reflectance spectra according to its nanostructure. Looking at the nanostructure of FlexBrite as shown in the SEM images in Fig. 1(b), Fig. 1S and Fig. 5(a), we propose that the high sensitivity for colorimetry is due to the combination of surface plasmonic resonance (SPR) of the periodic structures of nano-mushroom array and localized surface plasmonic resonance (LSPR) of the silver nanoparticles on the sidewall of nano-mushroom. Considering the complexity of the nanostructures of FlexBrite, the simulation is not trivial. To mimick the structure of nano-mushroom, we create a model with several variable geometric parameters, shown in Fig. 5(b), to resemble the real structure of nano-mushroom (Fig. 5(a)). The variable geometric parameters of nano-mushroom include the diameter $d_1$ of the sphere on the top, the diameter $d_2$ of the small spheres on the side walls, the conic angle θ of the polymer nanopillar, the thickness *t* of silver layer on the flat surface. From the SEM images of nano-mushroom, we can see there are a lot of particulate silver structures on the side walls of nano-mushrooms. We modelled the silver particles as small spheres on the sidewalls of nano-mushroom. To match the measured reflection peaks with the simulated reflection peaks at different liquid RI, the method of particle swarm optimization was used to find the optimal geometric parameters which allow the simulated peaks to be matched with measured peaks. Due to some irregularities of the real nanostructures as we can see in the SEM images in Fig. 1(b), Fig. 1S and Fig. 5(a), for the same FDTD model, it is impossible to get exactly the same peaks for simulation and measurement at different RI of liquid. In the optimization, we tried to minimize the difference between the simulated peaks and measured peaks all added together. Fig. 5(c-e) are some tentative geometries during the optimization with Fig. 5(e) as the optimal geometry we finally got. Fig. 4(b) is the simulated reflection spectra after optimization. The simulated peaks were plotted in Fig. 4(c) to compare with measured peaks. The peaks of reflection spectra in Fig. 4(b) match very well with the measured peaks in Fig. 4(a).

SERS characterization

To characterize FlexBrite on its surface-enhanced Raman spectroscopy (SERS) properties, we used 1,2-bis(4-pyridyl)ethylene(BPE) which will form a uniform monolayer on the silver surface after surface functionalization. For functionalization with BPE, FlexBrite was immersed in 100 μM BPE ethanolic solution for 5 h and then rinsed in pure ethanol, followed by blow drying with nitrogen to ensure a uniform single molecular layer adsorbed on the surfaces.[26] In ref [26] the BPE functionalization time is 24 hours. However, previously we did a kinetic study of SERS intensity versus functionalization time, shown as Fig. 4S in the supplementary information. From Fig. 4S, we found out the minimum functionalization time for stable and reliable SERS intensity is 5 hours. For Raman spectroscopy, we used the Renishaw Raman spectrometer with 20X objective lens and a He-Ne laser with 633 nm wavelength for excitation. The Raman enhancement factor (EF) was calculated by comparing the SERS signal intensity of BPE on FlexBrite with BPE bulk solution. The detailed procedures for EF calculation was included in the supplementary materials. The EF of FlexBrite in dry state which means without liquid on its surface, was calculated as $6.85 \times 10^7$. To verify the EF of FlexBrite we calculated, we used the commercial Klarite SERS substrate which is claimed to have the EF of $10^6$. We treated the Klarite substrate the same way as we treated FlexBrite with BPE incubation and measured the Raman signal. Fig.6 (a) shows the SERS comparison of FlexBrite with Klarite. With the same laser power and integration time for excitation, we found the SERS signal of FlexBrite is 60 to 80 times stronger than Klarite. For visual reason, we used higher excitation power of 1.23 mW for Klarite SERS and smooth silver while lower excitation power of 211 μW for FlexBrite SERS and plotted them on the same graph as Fig. 6(a) for comparison. To characterize the uniformity of SERS of FlexBrite, we used the mapping function of Renishaw Raman spectrometer, that is, a spatial 2D array of single Raman measurements at each point in the array. Fig. 6(b-d) show the result of SERS uniformity testing of FlexBrite with a 21×21 array with the period spacing of 0.5 mm for mapping. Fig. 6(b) is a series of SERS spectra of BPE at different locations in the mapping array. Fig. 6(c) is the averaged SERS spectra of all the measurements in the mapping array, with the cyan shadow as the standard deviation. In Fig. 6(d), the peak intensity at the wavenumber of 1607 cm$^{-1}$ was used to plot the distribution of SERS enhancement for the mapping array on FlexBrite. The variation of SERS intensity is below 15% for the area of 10 mm × 10 mm.

The previous measurement was done in dry state, which means without liquid on the surface. Considering the FlexBrite will present color change with liquid of different RI on the surface, we are interested in the SERS of FlexBrite with liquid on the surface, or wet state. Shown as Fig. 6(e), water was dropped on the surface of FlexBrite after BPE incubation for 5 hours as wet state. From top view, FlexBrite became green with water due to its colorimetric property. Then the 633 nm laser was shined and focused on the surface of FlexBrite for spectrum acquisition. With the same configuration we used for measurement in dry state, we found that the SERS signal of BPE for wet state FlexBrite is about 7 times stronger than SERS in dry state. Fig. 6(f) shows the comparison of dry state (RI = 1) SERS and wet state (RI = 1.333) SERS of FlexBrite. To investigate the phenomenon that wet-state SERS of FlexBrite is much stronger than dry-state SERS, we compared both the measurement and FDTD simulation of the reflection spectra of FlexBrite in dry and wet state, shown as Fig. 7. The solid curves in Fig. 7(a) are the measured reflection spectra for dry state (blue) and wet state (red) while dashed curves in Fig. 7(a) are the simulated reflection spectra for dry state (blue) and wet state (red). For both measured and simulated reflection spectra, we can see a valley close to the excitation wavelength of 633 nm for wet state while no valley was seen around 633 nm for dry state. This indicates that the nano-mushroom structure of FlexBrite is on resonance at 633 nm in wet state. The 7 times' enhancement of SERS in wet state is caused by the plasmonic resonance around the excitation wavelength of 633 nm. According to the electromagnetic theory of SERS, the SERS enhancement factor is proportional to $E^4$, the 4$^{th}$ power of electric field intensity.[27] Around the metallic nanostructures, only the locations where the electric field is very strong will have SERS effect and those locations are referred to as "hot spot". To verify the enhancement due to resonance, from the FDTD simulation model we took the cross-sections of electric field distribution around nano-mushroom at 633 nm in Fig. 7(b). In the FDTD model, the intensity of the electric field of incident wave is defined as unit 1. The electric field distribution in dry state is shown on the left and that in wet state is shown on the right with the same color bar as the scale of electric field intensity compared to the incident field. From Fig. 7(b) we can see that there are more "hot spot" around the nano-mushroom of FlexBrite in wet state than in dry state. This indicates that the nanomushroom is on resonance of 633 nm in wet state and explains the stronger SERS of FlexBrite in wet state.

Visible detection of biotin-streptavidin interaction by FlexBrite

Detection of biomolecular interaction such as antibody-antigen binding, DNA hybridization common and critical in biochemical assays. Extensive explorations have been done for easy and fast label-free detection of biomolecule interactions using the methods such as field effect transistor (FET) sensor[28], microcantilever sensor[22], localized surface plasmon resonance (LSPR)[29] and whispering gallery mode microcavities.[30] In this work, to demonstrate the label-free detection of biomolecular interaction with FlexBrite by both naked-eye colorimetry and SERS, we chose biotin–streptavidin interaction as a model system because they are well studied binding partners that interact with very high affinity and often used for proof-of-concept for detection of biomolecular interaction.[31, 32] We used the thiolated biotin to form a uniform monolayer on the silver surface of FlexBrite after 24 hours' incubation. The biotin-streptavidin functionalization was done by incubating the FlexBrite with biotin in streptavidin solution for 2 hours then rinsed with ethanol. Fig. 8(a) shows the schematic of biotin functionalization and biotin-streptavidin binding on the surface of FlexBrite. Prior to the measurement with colorimetry and SERS, we dropped water on the surface on FlexBrite just like how we characterized the FlexBrite in wet state. We measured the reflection images, spectra and SERS before and after the biotin functionalization and biotin-streptavidin binding respectively. Fig. 8(b) shows the colorimetric images of FlexBrite in water with biotin monolayer and after biotin-streptavidin binding. We can see that after formation of biotin monolayer, the color of FlexBrite changed a lot from green to green/yellowish. After biotin-streptavidin binding, the color changed a little towards the red, close to yellow. Fig. 8 (c) and (d) show the averaged RGB and HSV values for FlexBrite in water, with biotin and after biotin-streptavidin binding. The R value shows a clear trend of increasing after each step, indicating a red-shift, shown in Fig. 8(e). From the reflection spectra in Fig. 9(a), we can see the red-shift of reflection peak after each step. The reflection peak was shifted from 508 nm in water to 548 nm for biotin monolayer then to 575 nm after biotin-streptavidin binding. In addition to colorimetry, SERS was also used to interrogate the biotin-streptavidin binding. In Fig. 9(b), the SERS spectra of biotin-monolayer on FlexBrite were compared before and after binding with streptavidin. After biotin-streptavidin binding, the intensity of SERS was lower and fewer peaks showed up. The modification of the SERS spectrum implies a change in the secondary structure of the host biomolecule biotin. For example, The much less intense peak at 1287 cm$^{-1}$ in the presence of streptavidin implies the decrease of the amount of the β-sheet conformation.[33] Considering SERS is a near-field effect which decreases 10 to 12 fold with the distance from the analyte to the metal surface.[34] the weaker overall SERS signal after streptavidin binding may be due to the coverage of the 52.8 kDa protein streptavidin which blocks SERS signal from the nanomushroom.

SERS detection of narcotic drug methamphetamine in drinking water with handheld Raman spectrometer and FlexBrite.

The low cross-section of Raman scattering requires either highly-concentrated sample or highly-sensitive Raman spectrometer to get detectable signal. For this reason, the low-cost and convenient handheld Raman spectrometers are mostly used to identify highly-concentrated samples such as minerals due to their relatively lower sensitivity.[35] If the sensitivity of handheld Raman spectrometer can be greatly enhanced by SERS, then the fast and easy detection of trace amount of most organic chemicals can be implemented. To demonstrate that the sensitivity of handheld Raman spectrometer can be boosted by FlexBrite, we used the SERS property of FlexBrite to detect the narcotic drug methamphetamine in drinking water with a handheld Raman spectrometer B&WTek NanoRam handheld Raman spectrometer. The testing process was recorded as Video S3 in the supplementary materials. Fig. 10 showed several snapshots of Video S3 for the detection of methamphetamine. A small piece of FlexBrite was cut off from the 4 inch wafer and glued onto the inner wall of a glass vial used with NanoRam spectrometer, shown as Fig. 10(d). In Fig. 10(a), the methamphetamine was diluted in water with the concentration of 1mg/L (about 1 ppm). Then the diluted 1 ppm methamphetamine solution was sent by a pipette into a vial without FlexBrite for Raman testing with NanoRam spectrometer. The test result of vial without FlexBrite was shown in Fig. 10(c), in which no Raman peaks of methamphetamine was seen. The big hump of the spectrum is from the autofluorescence of the glass vial. Then the same 1 ppm methamphetamine solution was added into the vial with FlexBrite and tested with NanoRam spectrometer with the same configuration. Fig. 10(d) shows that the Raman peaks of methamphetamine were detected with FlexBrite in the vial. We proved that with FlexBrite the sensitivity of handheld Raman spectrometer can be greatly enhanced therefore can be used for easy and fast detection of trace amount of chemicals.

To find the limit of detection of methamphetamine with FlexBrite, we tested methamphetamine in aqueous solution with different concentrations. We put the data as Fig. 5S in the supplementary information. The lowest concentration of methamphetamine we could constantly detect is 0.5 mg/L. When the concentration of methamphetamine is high as 100 mg/L, the Raman signal is higher. The Raman signal does not vary too much in the concentration range of 1 mg/L to 10 mg/L. The explanation is that at high concentration much of the Raman signal comes from bulk solution itself so the Raman intensity is related to the concentration. But at lower concentration, the Raman signal mostly comes from SERS, which comes from molecules adsorbed on the FlexBrite surface so the Raman intensity does not change much with the concentration. 0.5 mg/L is the lowest concentration we can constantly detect Raman signal. When the concentration gets lower than 0.5 mg/L we can get the Raman signal occasionally but not always. As a result, we claim that 0.5 mg/L is the reliable limit of detection of methamphetamine.

Experiment section

Fabrication process
The nano-mushroom substrate was fabricated by nano-replication process. The cross-sectional schematic of the fabrication process was demonstrated in Fig. 6S in the supplementary materials. The original mold was nano-cup array on silicon oxide wafer fabricated by laser interference lithography followed by ion milling deep reactive ion etching.[20] Prior to replication, the mold was treated with silane for hydrophobicity.[36] The nano-pillar array was produced by molding the nano-cup array with Norland Optical Adhesive 61 ("NOA 61"). Then 9 nm of Titanium and 90 nm of silver were deposited onto the nano-pillar array by e-beam evaporation. The Titanium layer is to improve the adhesion. After the metal deposition, the nano-mushroom structure will be formed, shown as Fig. 6S. However, to this point, the nano-mushroom surface is very hydrophobic, which will prevent the aqueous solution from contacting the surface and the adsorption of target molecules. To make the surface hydrophilic, we deposited a very thin layer of silicon oxide $SiO_2$ onto the nano-mushroom surface by physical vapor deposition (PVD) for 5 seconds. We used the K. J. Lesker PVD 75 system for the thin $SiO_2$ deposition. We inflow Argon to 1.5mTorr after vacuum the chamber down to 3e-6 psi. Then 300 W plasma was ignited with the 3" PVD gun for the deposition at room temperature. The actual thickness of $SiO_2$ layer is too thin to measure directly. But by ellipsometry we measured the deposition rate for thicker $SiO_2$ with longer deposition time. The deposition rate can fluctuate. On average, the deposition rate was found to be 1 angstrom per second. We barely deposited for only 5 seconds for the thin layer. As a result, the thickness of silicon oxide is estimated to be about 5 angstroms. BPE SERS comparison of FlexBrite before and after $SiO_2$ deposition for 5 seconds is shown as Fig. 7S in the supplementary information. It shows comparable enhancement before and after $SiO_2$ deposition.

Colorimetry measurement

The colorimetry measurement was taken with Olympus BX51 upright fluorescence microscope equipped with DP50 digital camera and a mercury lamp. We keep the same camera setting for all the images. Exposure time of 30 ms. RGB gain was set to be 1. The microscope was set at bright field mode without applying any filter. The light was focused onto the sample with 20X (NA = 0.45) objective lens. The reflectance spectra were collected by the sample microscopy setup with USB2000+ Fiber Optic Spectrometer by Ocean Optics. The detection wavelength range is from 200 to 1100 nm and the spectral resolution is up to 0.3 nm FWHM.

FDTD simulation

The numerical simulations of optical characteristic were studied by using three dimensional finite-difference time-domain (3D-FDTD) method with FDTD software package from Lumerical Solutions, Inc. The x-axis polarized electromagnetic wave was set to propagate normal to the substrate (-z direction) for reflection, transmission, and near-field simulations. Perfect matching layer (PML) was applied to the boundary conditions in z axis to eliminate any interference from the boundaries. In addition, PML and periodic boundary were applied at x and y axes for simulating single and array nanostructure respectively.

SERS measurement

Renishaw PL/Raman micro-spectroscope system was applied for SERS signal measurements. 633 nm He-Ne lasers with the power of 5 mW are used as excitation light source. 20X objective lens (NA = 0.45) was used to focus/collect incident light and Raman signal onto/from the surface of FlexBrite. The range of collected wavenumber was from 200 to 2000 $cm^{-1}$. The laser power was attenuated to 1.23 mW and 211 µW with neutral density filters. The methamphetamine solution, 1.0 mg/mL in methanol, ampule of 1 mL certified reference material and 1,2-bis(4-pyridyl)ethylene(BPE) were purchased from Sigma Aldrich.

Handheld Raman spectrometer

We used B&WTek NanoRam® handheld Raman spectrometer along with its software NanoRam® ID to measure the SERS spectrum of methamphetamine solution. The vial adapter and a glass vial were used to contain the methamphetamine solution. The excitation laser wavelength is 785 nm; laser power is 5mW. The spectral range is 176$cm^{-1}$ to 2900$cm^{-1}$. The total data acquisition, averaging, background acquisition and post processing took about 2 minutes. More details about B&WTek NanoRam® can be found on the B&WTek website: http://bwtek.com/products/nanoram/

Protocol for biotin-streptavidin binding

For functionaliztion of monolayer of biotin: 1. Rinse the FlexBrite with isopropanol followed by water then by isopropanol again. 2. Incubate the FlexBrite in thiolated-biotin solution (1mM) for 24 hours at 37 °C. 3. Rinse the substrate with ethanol twice. For biotin-streptavidin binding: 4. Incubate the biotin-functionalized FlexBrite in streptavidin (1 µg/ml) for 2 hours. Thiolated biotin was purchased from Nanocs Inc and stock solution was prepared with ethanol dilution. Streptavidin was purchased from Life technology and stock solution is prepared with 1XPBS.

Conclusion

In this work, we demonstrated a novel dual-mode plasmonic and SERS substrate with large-area uniformity, low cost and mass-producibility. The plasmonic colorimetry was characterized with the sensitivity of 611 nm/RIU, which enabled the naked-eye detection of biotin-streptavidin interaction. The SERS enhancement factor was characterized as $6.85 \times 10^7$ in dry state and $4.81 \times 10^8$ in wet state. The SERS property significantly enhanced the sensitivity of the handheld Raman spectrometer to enable the easy and fast detection of trace amount of narcotic drug methamphetamine in drinking water.

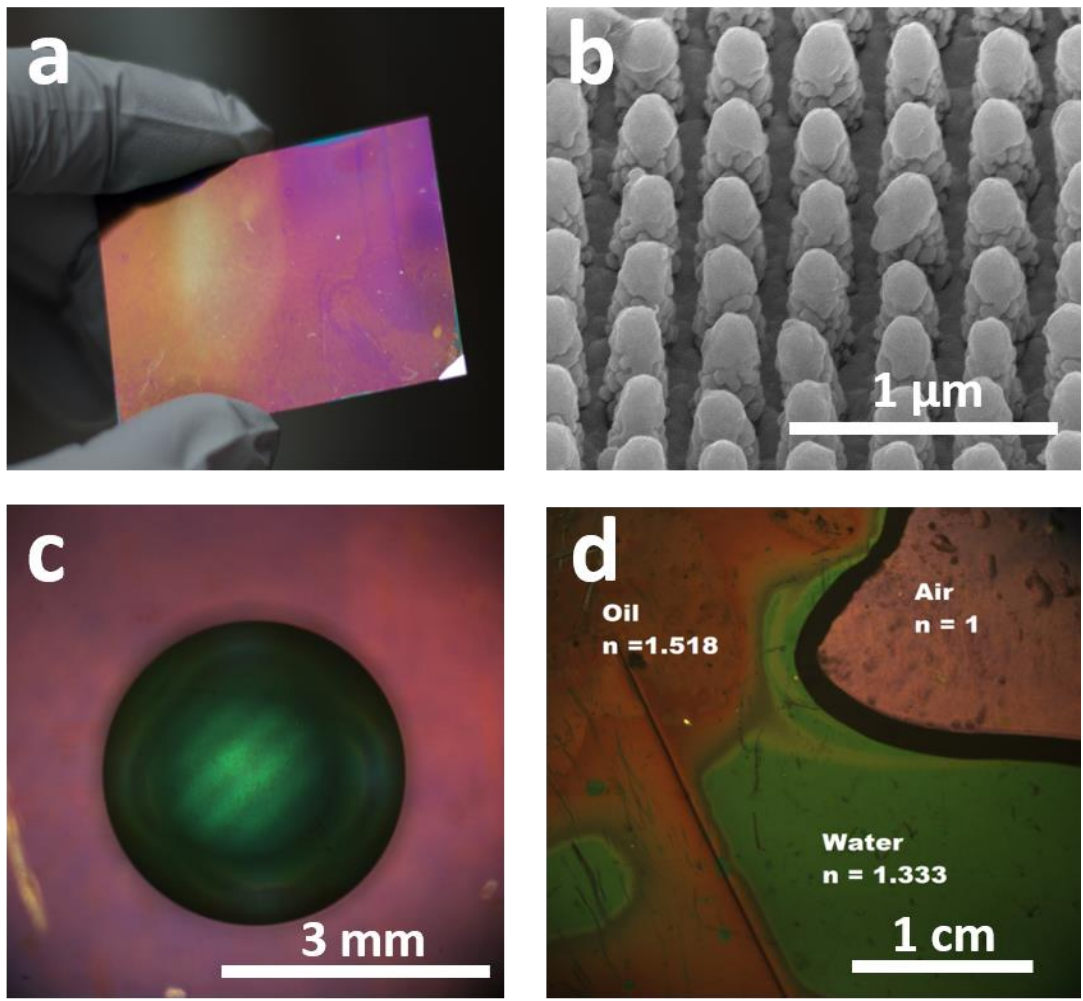

Fig. 1. Overview of FlexBrite substrate. (a) Photograph of one piece of FlexBrite. (b) Scanning electron microscopy (SEM) image of surface of FlexBrite. (c) A water droplet on the surface of FlexBrite shows green color. (d) Water (n = 1) and cedar wood oil (n = 1.518) on the surface of FlexBrite show green and red color respectively.

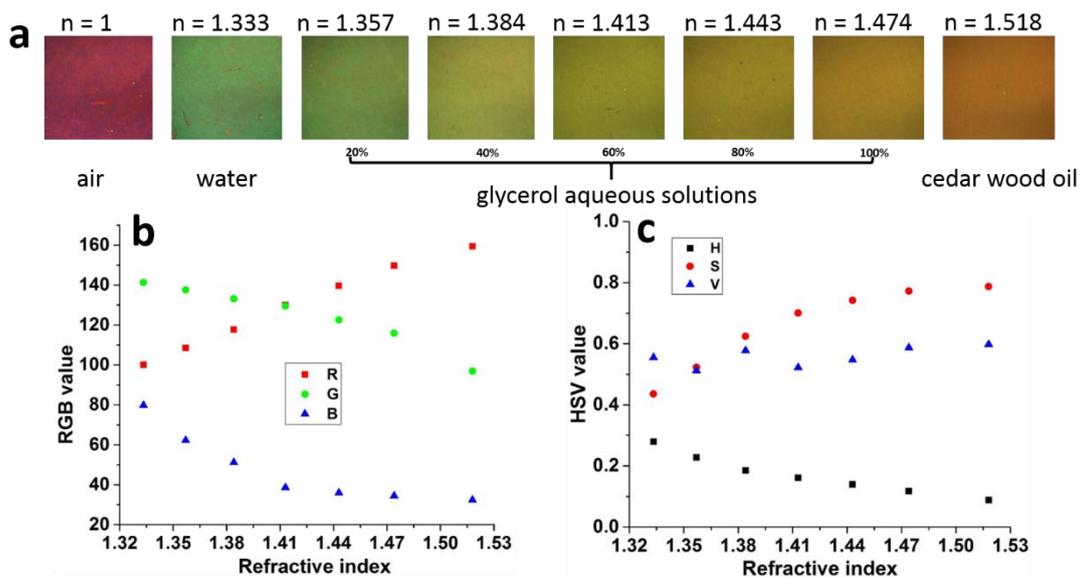

Fig. 2. Plasmonic colorimetry with FlexBrite. (a) Images showing different colors with different refractive index (RI) of liquid on the surface. (b) Averaged RGB values with different RI of liquid. (c) Averaged HSV values with different n of liquid.

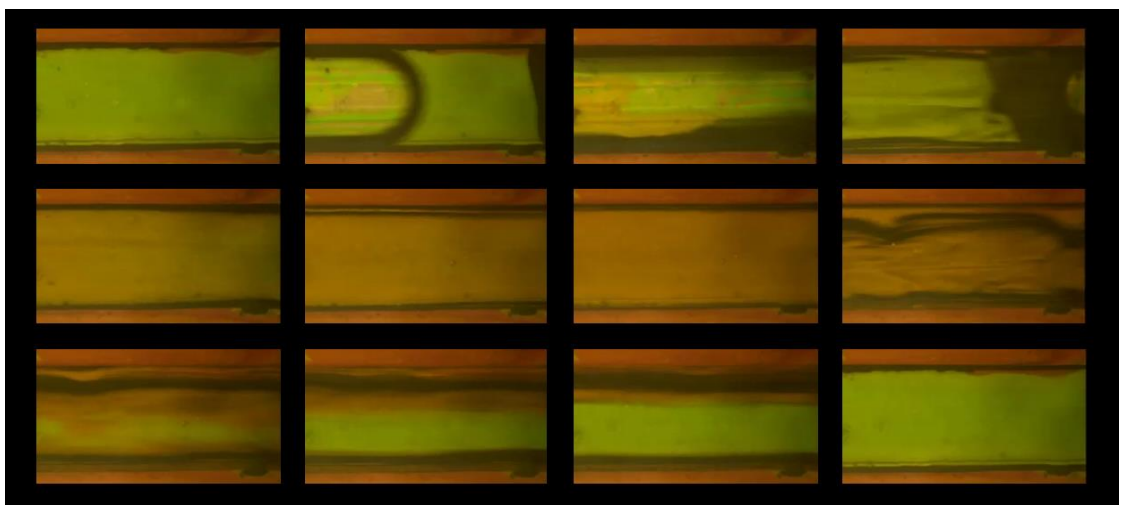

Fig. 3. A series of video snapshots to demonstrate the reusability of FlexBrite in microfluids. In a microfluidic channel (width = 500 µm) with FlexBrite in the bottom, water is first replaced with glycerol then glycerol is replaced with water again.

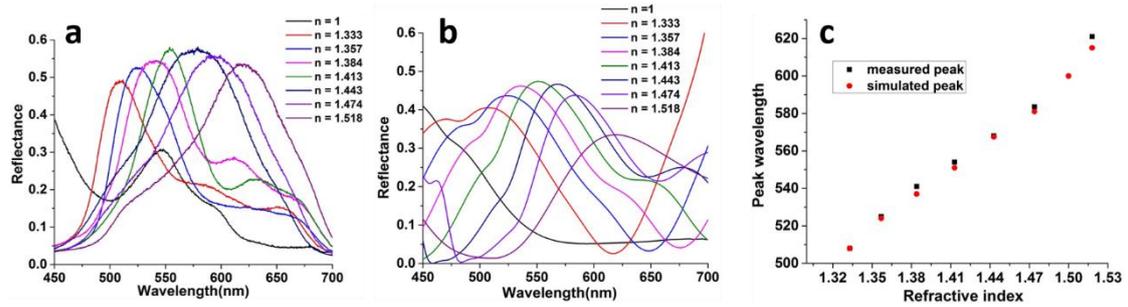

Fig. 4. Spectroscopic characterization of FlexBrite. (a) Measured reflection spectra of FlexBrite with different RI of liquid. (b) Simulated reflection spectra of FlexBrite with different RI of liquid. (c) Measured (black) and simulated (red) reflection peaks with different n of liquid. Characterized sensitivity: 611 nm/RIU.

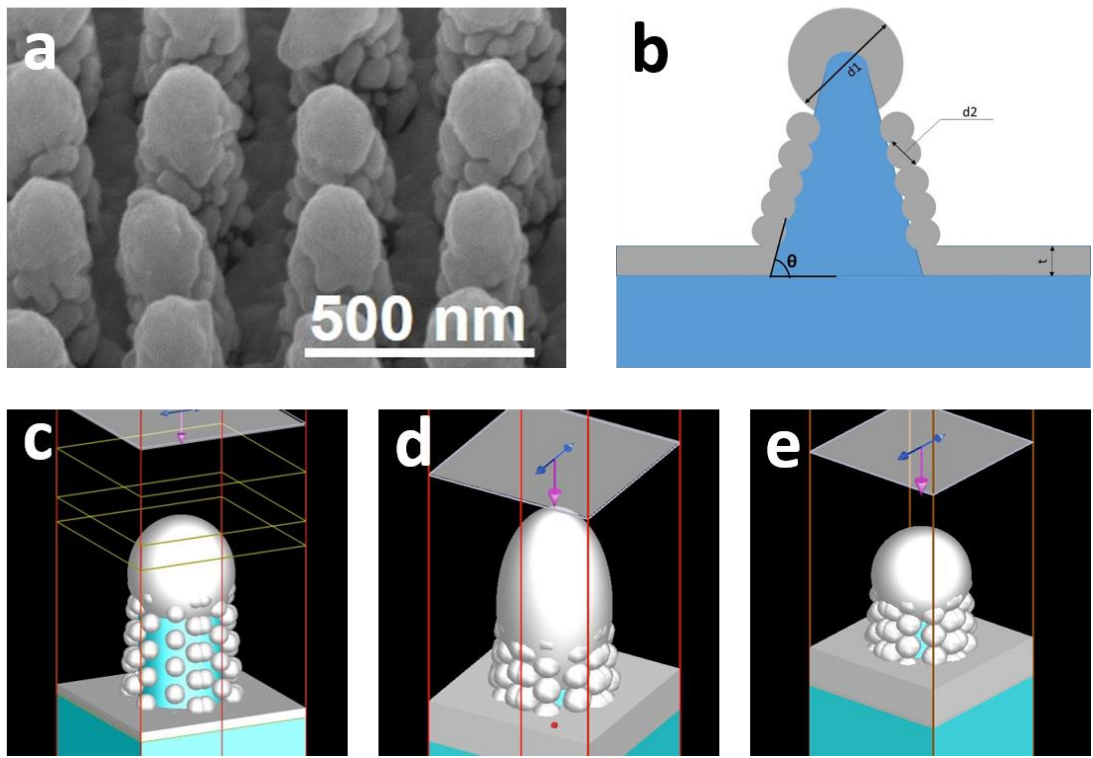

Fig. 5. Optimization of FDTD simulation for reflection peaks of FlexBrite with different RI of liquid. The simulated structure parameters such as mushroom width, mushroom height, and size of the silver nanoparticles were calibrated using particle swarm optimization, so that the simulated reflection spectrum matched the peak and trough locations of the experimental data. (a) SEM image of nano-mushroom structure on the surface of FlexBrite. (b) Schematic of the FDTD model of nano-mushroom and the parameters used in optimization. (c-d) Snapshots of FDTD models with different parameters in optimization. (d) The optimal geometry with the simulated reflection peaks closest to the measured reflection peaks.

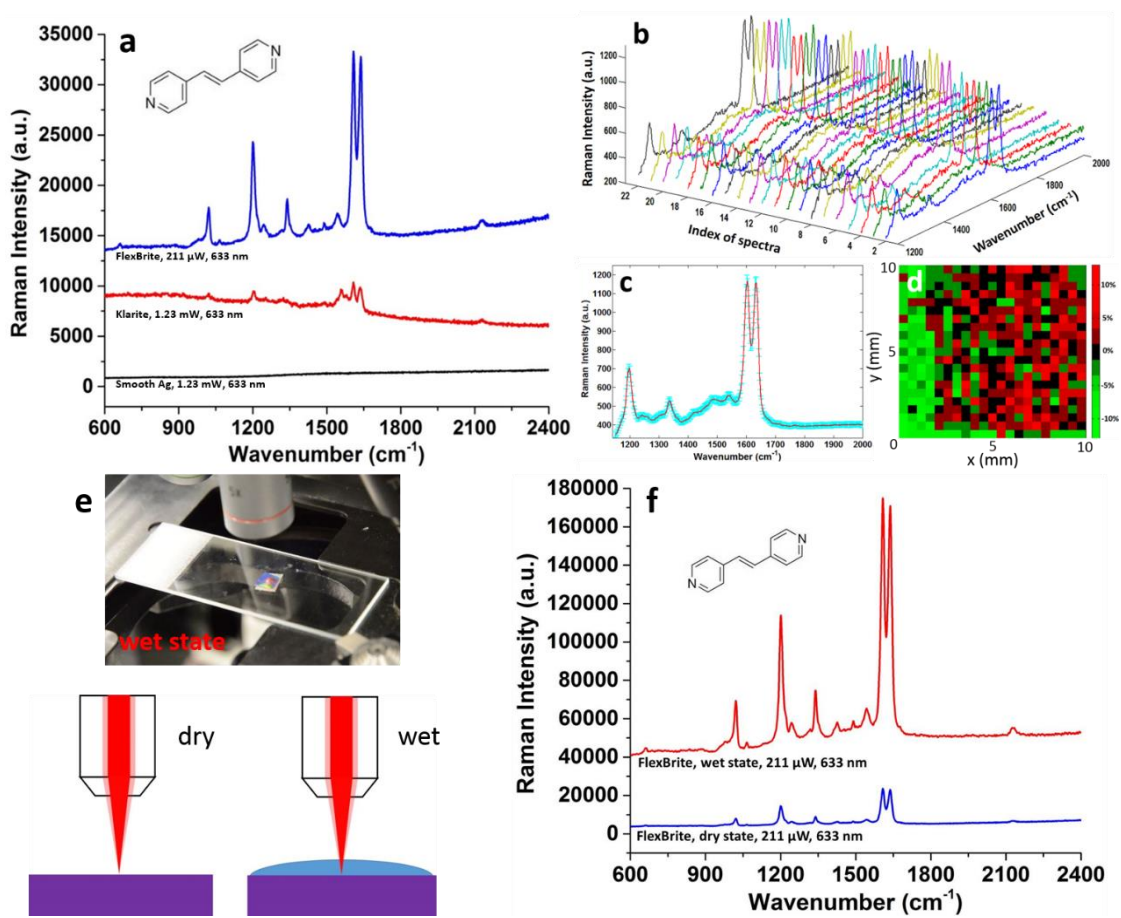

Fig.6. Surface-enhanced Raman spectroscopy (SERS) characterization of FlexBrite using 1,2-bis(4-pyridyl)ethylene (BPE)as the target molecule. (a) SERS comparison of FlexBrite, Klarite and smooth silver. (b) A series of SERS spectra of BPE at different spots on FlexBrite. (c) Averaged SERS (red curve) spectrum and standard deviation (cyan shadow) of BPE on FlexBrite. (d) Uniformity testing of SERS of BPE on FlexBrite by Raman mapping. (e) Illustration of dry state SERS and wet state SERS on FlexBrite. (f) Comparison of dry state SERS and wet state SERS of BPE on FlexBrite.

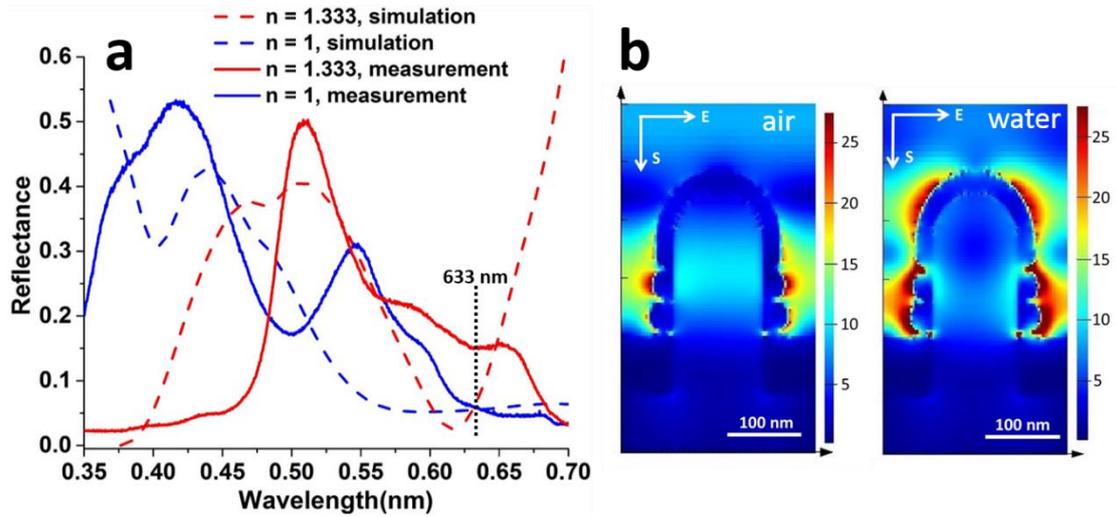

Fig. 7. Explanation of stronger SERS of FlexBrite in wet state. (a) Measured (solid curves) and simulated (dashed curves) of FlexBrite in dry state (blue curves) and wet state (red curves). Both measured and simulated reflection spectra show a dip around 633 nm, the excitation laser wavelength. This indicates the resonance at 633 nm. (b) Electric field distribution of the simulated nano-mushroom in dry state (left, RI = 1) and wet state (right, RI = 1.33).

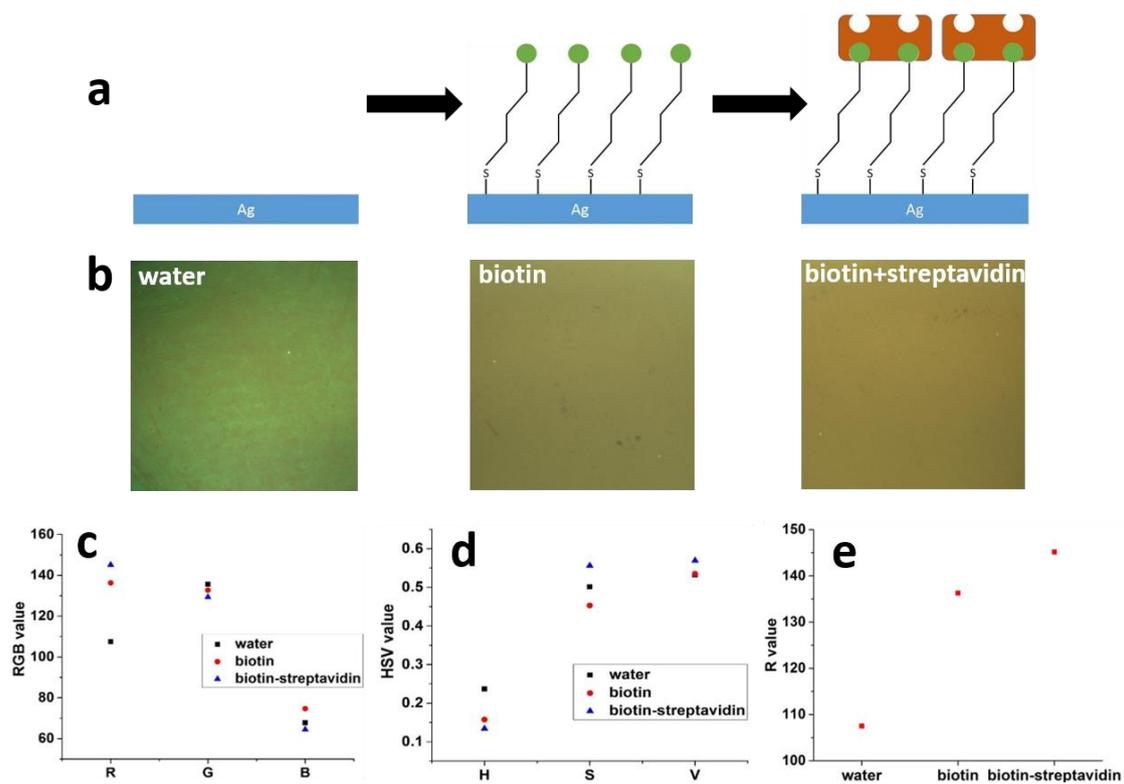

Fig. 8. Detection of antibody-antigen reaction (biotin-streptavidin binding) with plasmonic colorimetry of FlexBrite. (a)The process of functionalization of thiolated biotin on the surface of FlexBrite and conjugation with streptavidin. (b) Colors of FlexBrite in wet state with water, after functionalization of thiolated biotin and after binding biotin with streptavidin. (c) Averaged RGB values of FlexBrite with water, biotin and biotin-streptavidin. (d) Averaged HSV values of FlexBrite with water, biotin and biotin-streptavidin. (e) Averaged R value of FlexBrite with water, biotin and biotin-streptavidin.

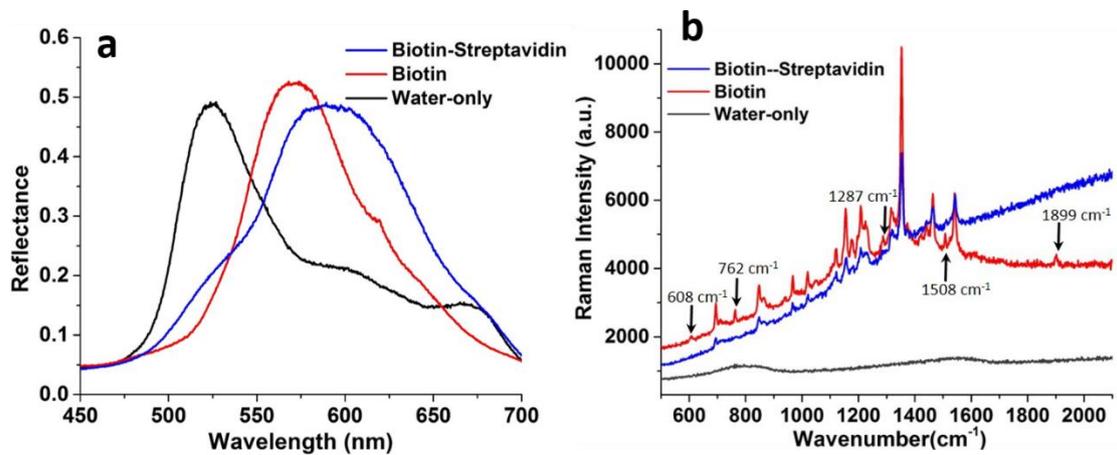

Fig. 9. (a) Reflection spectra of FlexBrite with water, biotin and biotin-streptavidin conjugate on the surface of FlexBrite. (b) SERS spectra of FlexBrite with water, biotin and biotin-streptavidin conjugate on the surface of FlexBrite.

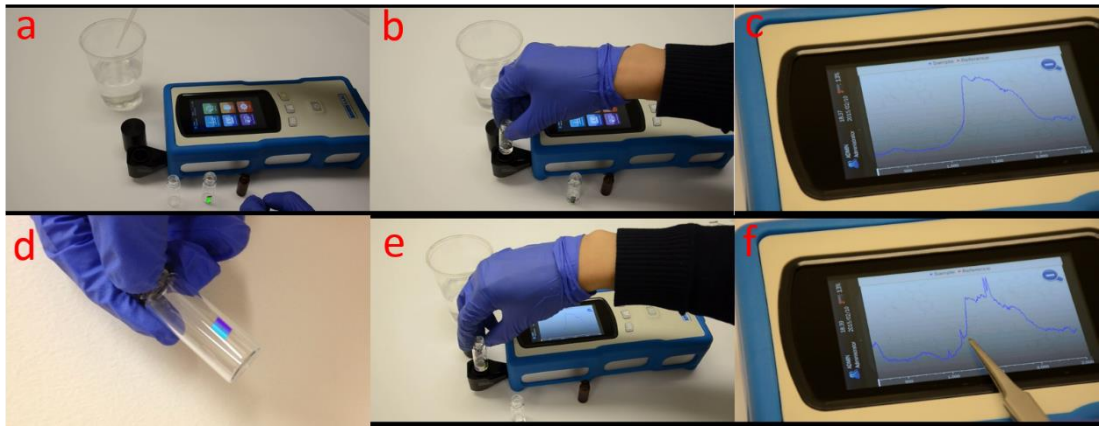

Fig. 10. Snapshots of a video to demonstrate the improved sensitivity of B&WTek NanoRam handheld Raman spectrometer with FlexBrite. (a) Diluting methamphetamine in water with the concentration of 1 mg/L (1 ppm). (b) Taking 1 ppm methamphetamine solution into the glass vial without FlexBrite for measurement with NanoRam spectrometer. (c) No Raman peaks were observed for 1 ppm methamphetamine solution without FlexBrite. (d) The glass vial with one piece of FlexBrite attached on the inner wall, facing outside. (e) Taking 1 ppm methamphetamine solution into the glass vial with FlexBrite for measurement with NanoRam spectrometer. (f) Raman peaks of methamphetamine were observed for 1 ppm methamphetamine solution with FlexBrite.

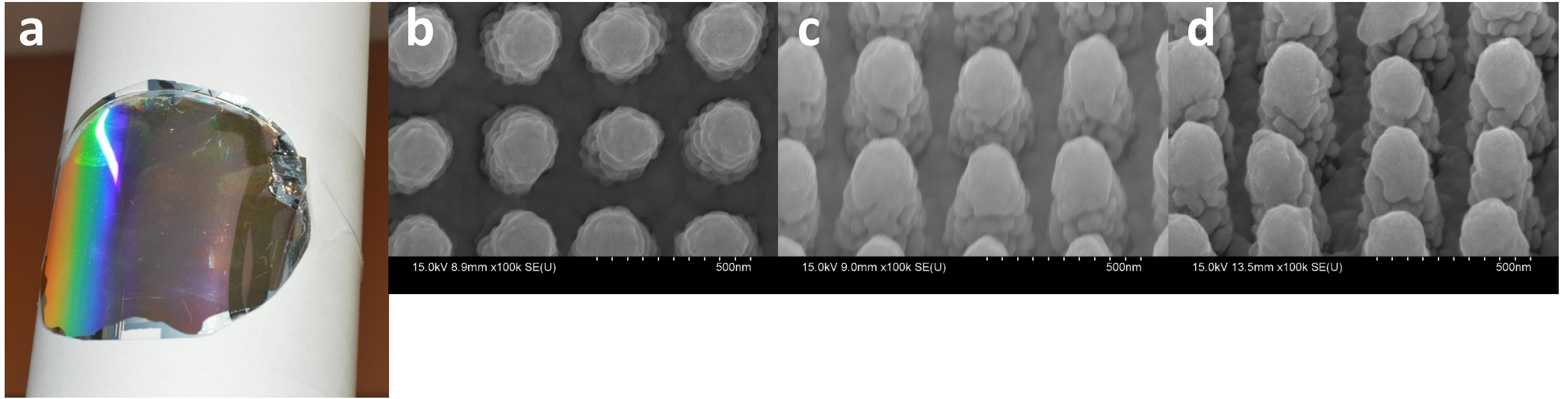

Fig. 1S. (a) A full-view photograph of one piece of FlexBrite with the diameter of 4 inches, wrapped around a mailing tube. (b) Top-view SEM of nanomushroom on FlexBrite. (c) 30º tilted view of nanomushroom on FlexBrite. (d) 45º tilted view of nanomushroom on FlexBrite.

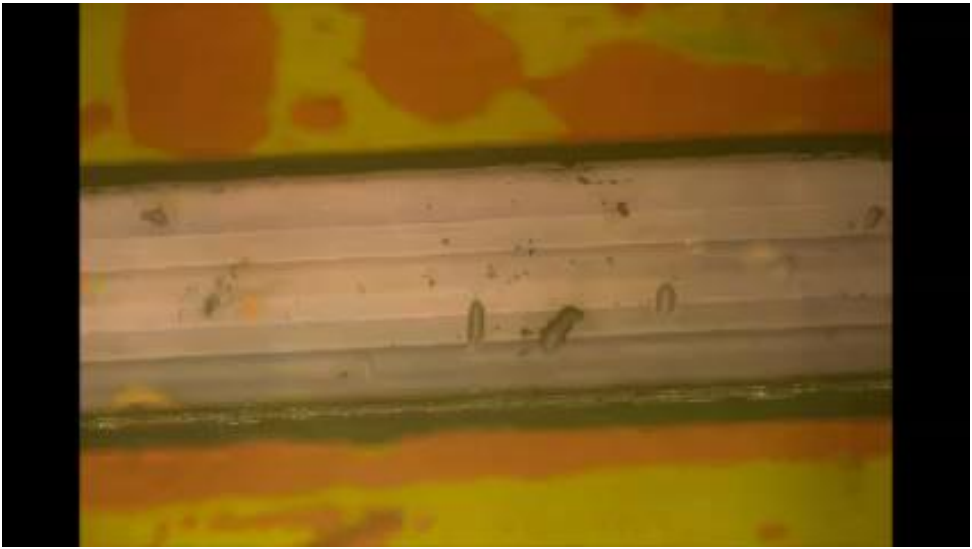

Video S1. Color change of FlexBrite in a microfluid channel when the refractive index of fluid is gradually increased from n = 1 (water) to n = 1.47(100% glycerol) , starting from air.

Link to Video S1:
http://v.youku.com/v_show/id_XMTQ0MTY3NDcwNA==.html

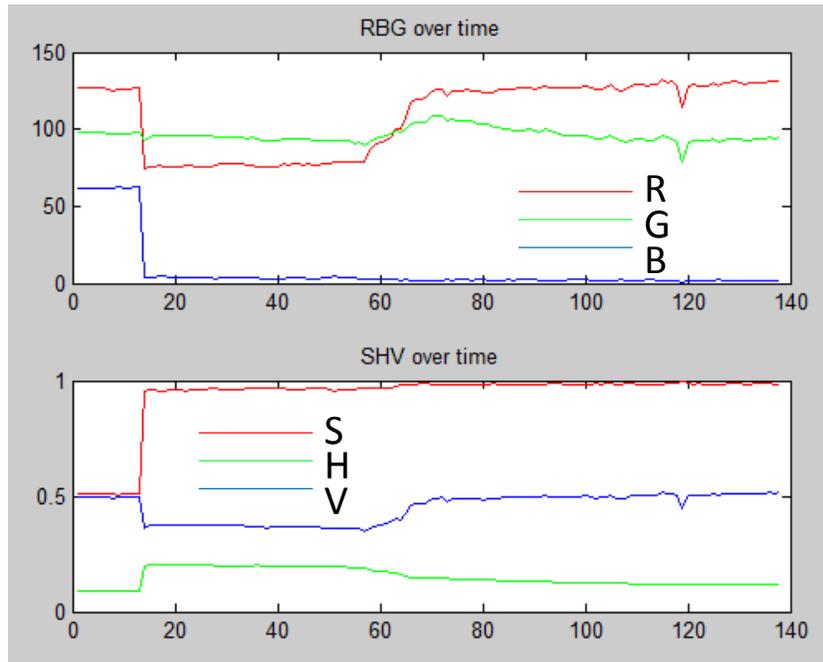

Fig. 2S. RGB (upper graph) and HSV (lower graph) values change of FlexBrite with time for the gradient increase of refractive index of fluid.

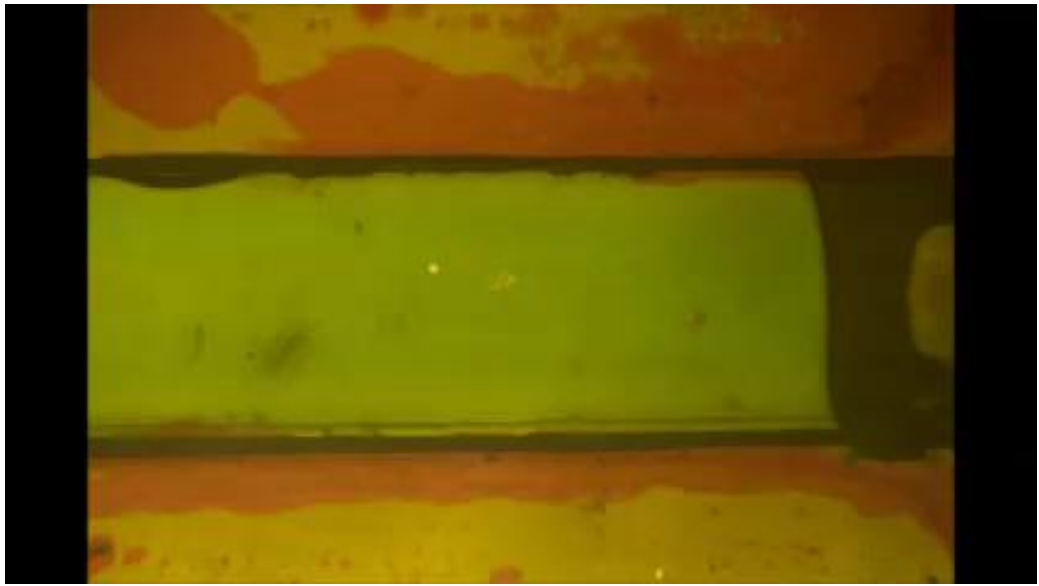

Video S2. Proof of reusability of FlexBrite. Observation of color change by alternatively replacing the fluid in microfluid channel with water and glycerol solution.

Link to Video S2:
http://v.youku.com/v_show/id_XMTQ0MTY3NTIwMA==.html

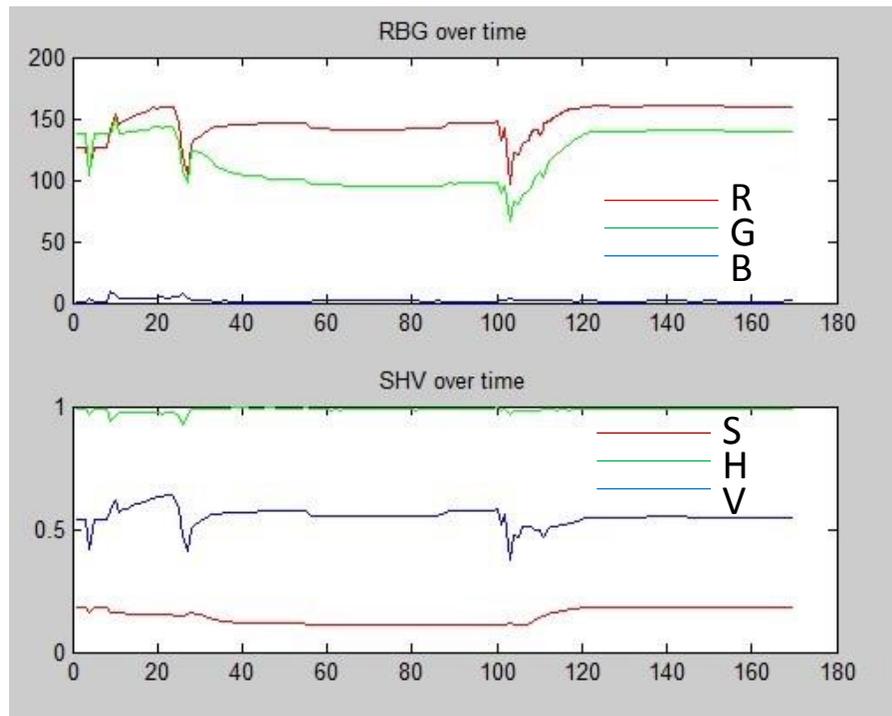

Fig. 3S. RGB (upper graph) and HSV (lower graph) values change of FlexBrite with time when alternatively replacing the liquid with water and glycerol.

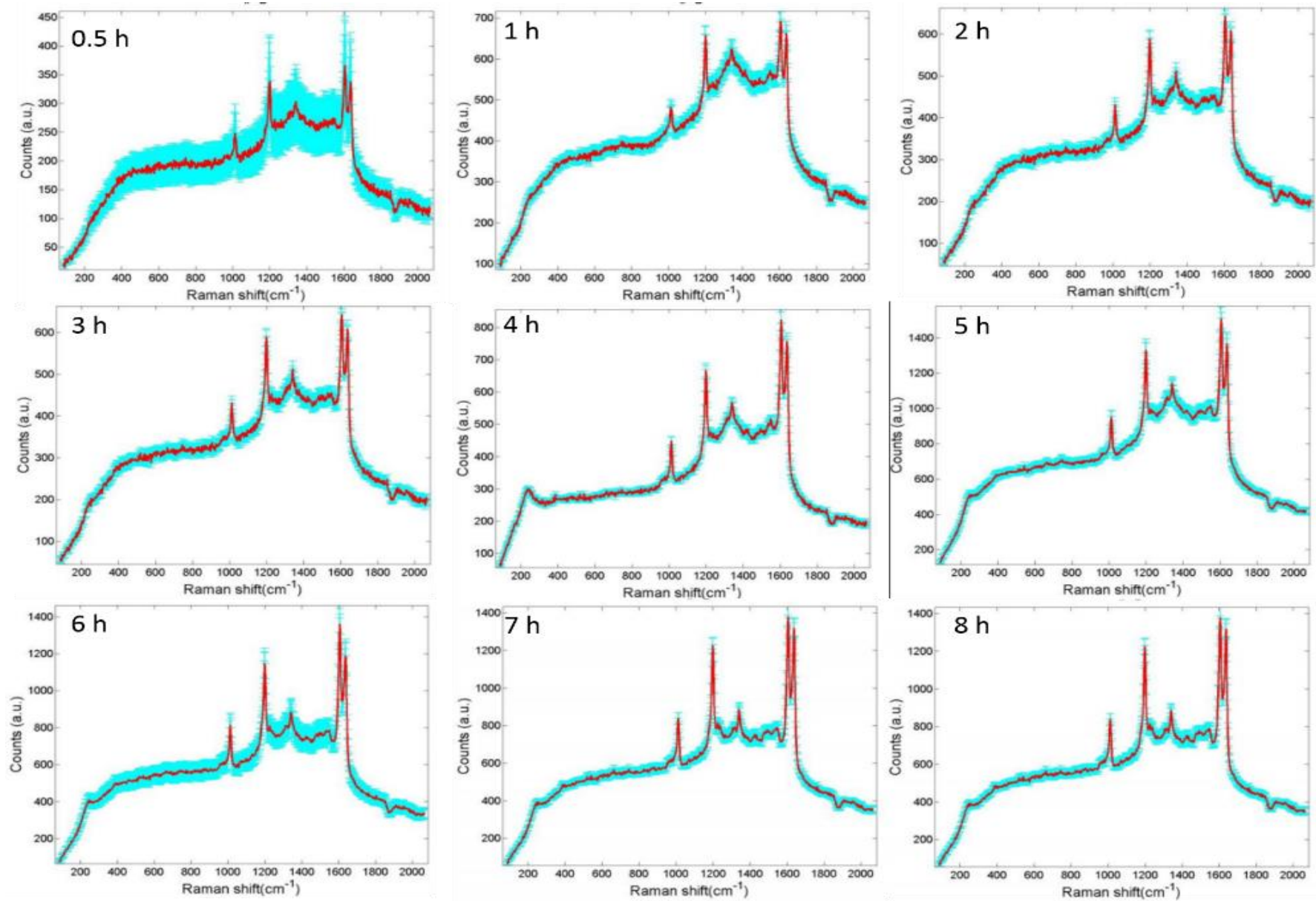

Fig. 4S. Kinetic study of BPE incubation time from 0.5 hour to 8 hours. After 5 hours' incubation the SERS signal become stable.

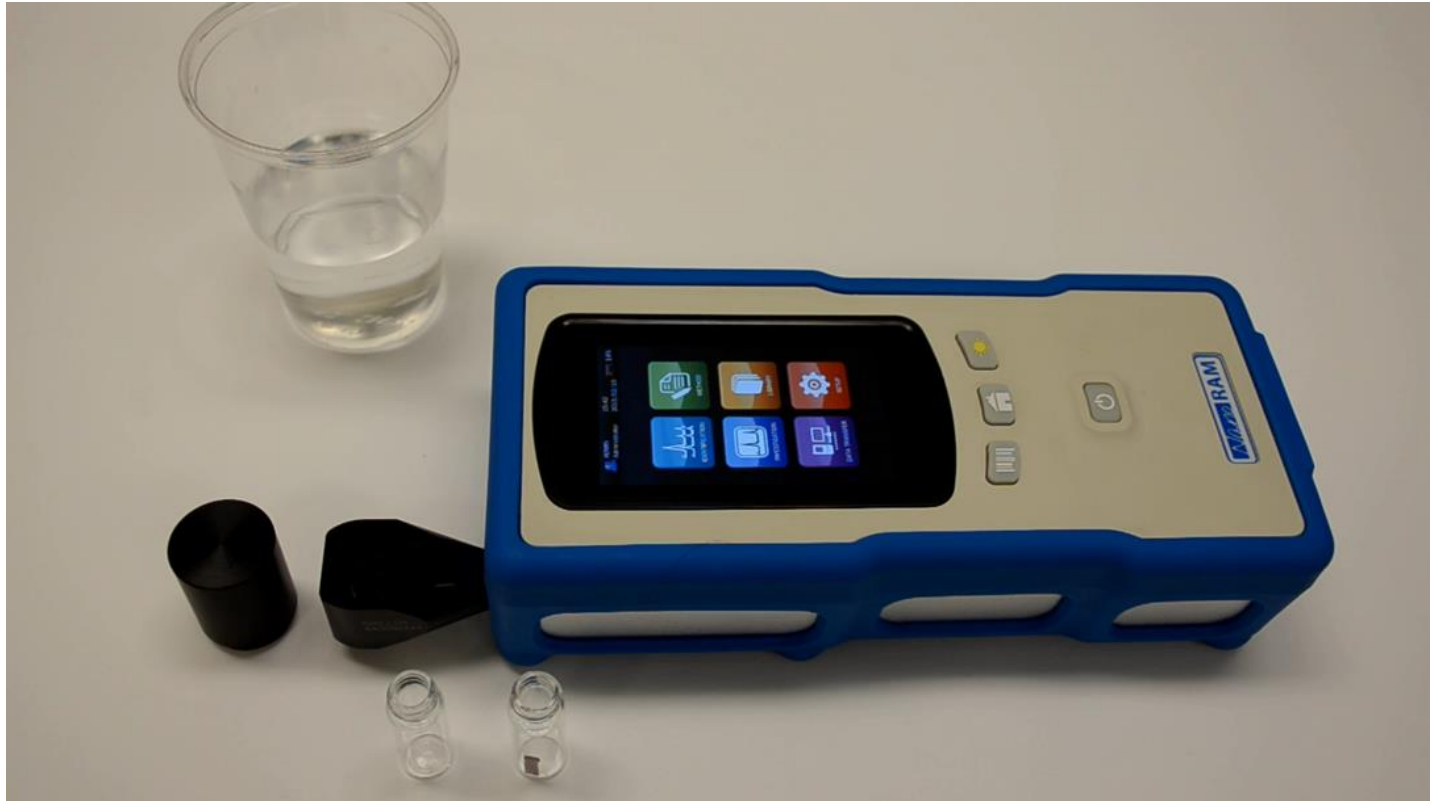

Video S3. Video to show the detection of methamphetamine in drinking water with B&Wtek NanoRam handheld spectrometer and FlexBrite.

Link to Video S3: http://v.youku.com/v_show/id_XMTQ0MTY3NTQ2OA==.html

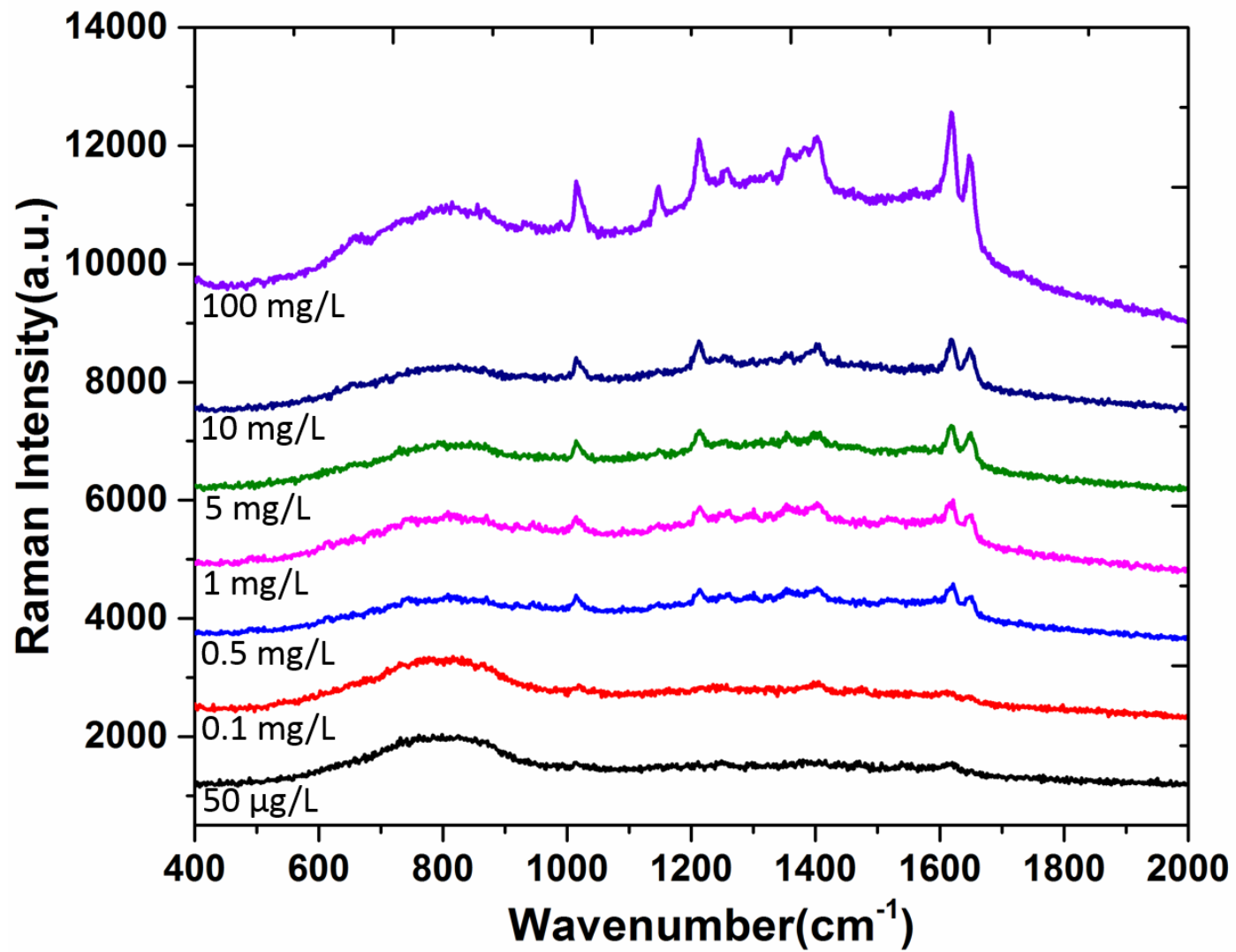

Fig. 5S . SERS of different dilutions of methamphetamine in aqueous solution. Limit of detection is 0.5 mg/L.

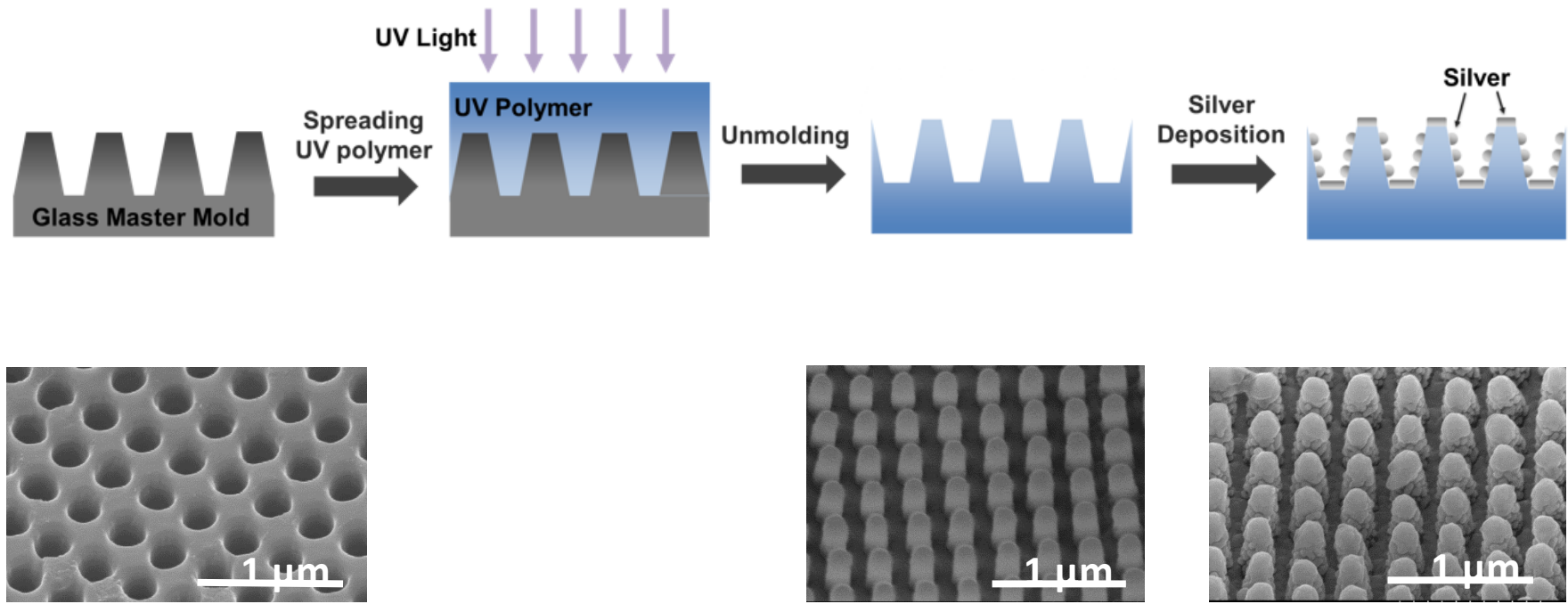

Fig. 6S. Fabrication process of FlexBrite. The glass master mold is made of silica. The nanohole array on silica was created by laser interference lithography followed by deep reactive ion etching. Firstly a 4 inch diameter (also performed with a 6 inch diameter) silicon dioxide wafer is coated with an average 0.45 µm thick photoresist and then exposed by 413 nm wavelength laser interference illumination with a dose of ~40 mJ cm$^{-2}$. After the photoresist development, the wafer is covered by a uniform array of nanoscale hollow photoresist mask of 150 nm in diameter and 350 nm in spacing distance. Secondly the wafer is subject to ion milling deep reactive ion etching by using a highly directional Bosch process. The unprotected circular area was etched down for 500 nm to form nanohole array. The last step is 90 nm Ag deposition after 9 nm Ti deposition as adhesion layer.

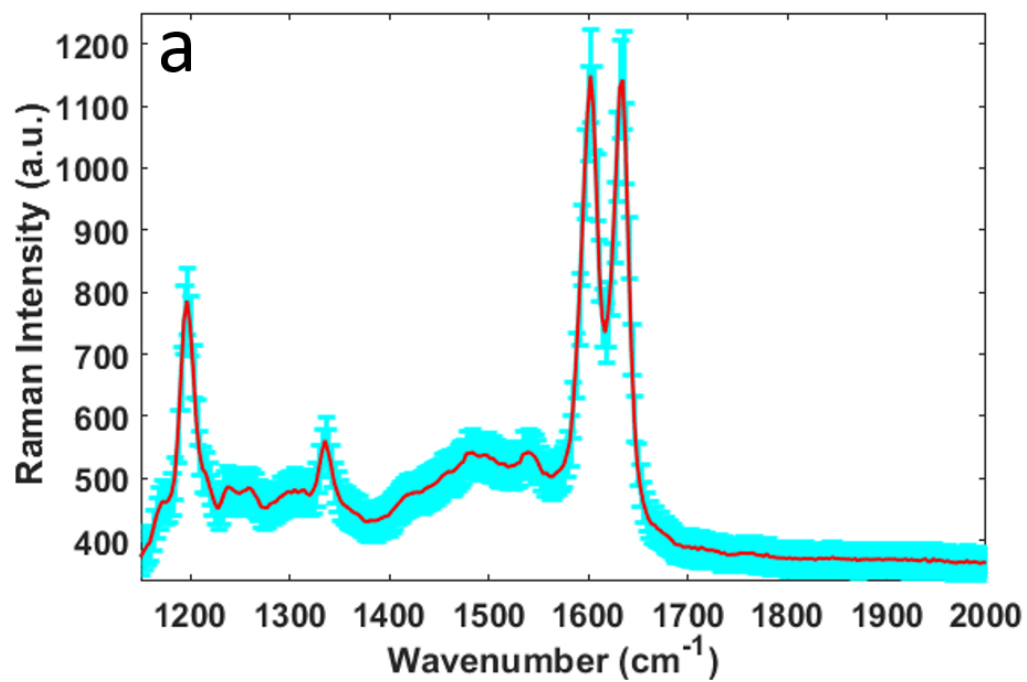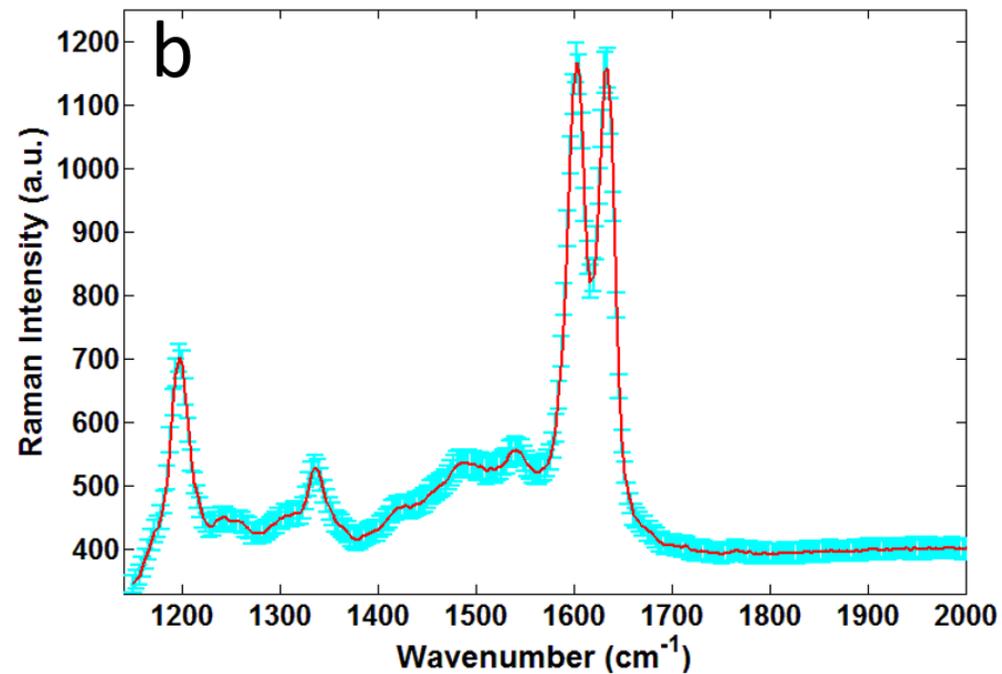

Fig. 7S. SERS comparison of FlexBrite before (a) and after (b) SiO$_2$ deposition for 5 seconds with BPE incubation for 5 hours.

Detail calculation of SERS enhancement factor (EF)

The average of SERS enhancement factor (EF) from FlexBrite sensor can be calculated as the following equation:

$$EF = \frac{I_{SERS}}{I_{Raman}} \times \frac{N_{Raman}}{N_{SERS}} \times \frac{P_{Raman}}{P_{SERS}} \times \frac{T_{Raman}}{T_{SERS}}$$

where $I_{SERS}$ and $I_{Raman}$ are integrated scattered intensities of Raman signal from FlexBrite sensor and BPE bulk solution. $N_{SERS}$ and $N_{Raman}$ are the number of molecules being probed on FlexBrite sensor and in the BPE bulk solution. $P_{SERS}$ and $P_{Raman}$ are the power intensity of excitation laser applied onto FlexBrite sensor and BPE bulk solution. $T_{SERS}$ and $T_{Raman}$ are the acquisition time when measuring Raman signal on the FlexBrite sensor and in the BPE bulk solution. $I_{SERS}$ and $I_{Raman}$ were measured with the integrated peak intensity at wavenumber 1607 cm$^{-1}$. The integrated Raman intensity of FlexBrite sensor and BPE bulk solution substrate were 235359.05 and 4585.83. For the calculation of probed molecule number, the surface area of top Ag layer, bottom Ag nanoparticle, and sidewall Ag nanoparticles need to be considered. We calculated the active Raman enhanced surface area as the total surface area of silver covered on the device. The model we applied here was the one used in FDTD simulation. The laser spot size and focal length after 20X objective lens were measured as 6.06 µm and 3 mm. The surface area of single BPE molecule is 30 Å$^2$.[S2] The concentration of BPE bulk solution was 100 mM. As a result, the $N_{SERS}$ and $N_{Raman}$ can be calculated as 1.59×10$^8$ and 5.22×10$^{12}$. The laser power intensities applied on FlexBrite sensor and BPE bulk solution were 211 µW and 1.23 mW. The acquisition time were 10 and 30 seconds for Raman measurements on FlexBrite sensor and BPE bulk solution. Finally, by applying all the parameters and EF equation above the SERS EF of FlexBrite sensor under solution-based environment can be estimated as 7.26×10$^7$.